\documentclass[prl,aps,showpacs,unsortedaddress,superscriptaddress,twocolumn]{revtex4}
\usepackage{graphicx}

\begin{document}

\title{Disentangling phonons from spins in ion-trap-based quantum spin simulators}

\author{C.-C. Joseph Wang and James Freericks}
\email{joseph@physics.georgetown.edu}

\affiliation{Department of Physics, Georgetown University, 37th and O Sts. NW, Washington, DC 20057, USA}

\date{\today}


\def\bx{{\bf x}}
\def\bk{{\bf k}}
\def\br{{\bf r}}
\def\bu{{\bf u}}
\def\half{\frac{1}{2}}
\def\args{(\bx,t)}

\begin{abstract}
We compute how phonon creation affects the fidelity of the quantum spin dynamics in trapped ion simulators.
A rigorous treatment of the quantum dynamics is made by employing an exact operator factorization of the evolution operator.
Although it is often assumed that phonon creation modifies the dynamics of the spin evolution, for an Ising spin-spin interaction in an external magnetic field, phonons have \textit{no effect} on the probabilities of spin product states measured
in the direction of the Ising model axis.
Phonons play a much more important role in influencing the effective spin dynamics
for Heisenberg or XY model spin simulators or for other observables, like witness operators in the Ising model.

\end{abstract}

\maketitle
In the early 1980's, Richard Feynman proposed that one could used controlled quantum-mechanical systems to simulate the many-body problem~\cite{feynman}. In recent years, there has been significant success in trying to achieve this goal~\cite{emulators}. We focus here on one platform for performing analog quantum computation, the simulation of interacting quantum spins in a linear Paul trap with a finite number of ions~\cite{gate,three-ion,Kim,A.Friedenauer}.  In these systems, the clock states of the ions are the pseudospin states, which can be manipulated independently by a spin-dependent force driven by laser beams. The lasers couple the pseudospin states to the lattice vibrations of the trapped ions, which leads to effective spin-spin interactions when the phonon degrees of freedom are integrated out~\cite{spin-spin-interactions,Duan,Kim}.  Recently, ion traps have been used to simulate the transverse-field Ising model with two~\cite{A.Friedenauer} and three~\cite{three-ion} ions in a trap.  Experiments are usually run in the Lamb-Dicke regime where spontaneous phonon creation is low, but it is generally believed that phonon creation, especially as one tunes the driving force close to resonance with one of the vibrational mode frequencies, will act to dephase and decohere the quantum dynamics leading to errors in the analog simulators that will build up as a function of time. Surprisingly, for the simplest class of emulators (namely the transverse-field Ising model) one can show that there is \textit{no dephasing or decoherence by the phonons to lowest order in the Lamb-Dicke parameter} if one measures the probabilities of product states of spins where the spin quantization axis coincides with the Ising model axis. For more complicated models, like the XY or Heisenberg models, phonon dephasing and decoherence cannot be removed, and hence this will create more stringent restrictions on the fidelities of quantum simulators for those models when simulated in ion traps. Moreover, when measuring entanglement witness operators, even in the transverse field Ising model, the phonons should affect the results.

Quantum spin models are of interest for a number of reasons.  Currently, there is a significant focus on the physics of frustration~\cite{Frustration}, and how that can lead to the formation of a so-called spin liquid~\cite{spinliquid}. These spin liquid states are complicated quantum many-body systems that exhibit significant entanglement  of their wavefunctions, and could also exhibit emergent phenomena within their low-energy excitation spectra. Conventional computation fails to describe these systems well.  Exact diagonalization studies are limited to small size lattices, while quantum Monte Carlo simulations suffer from the sign problem and cannot reach low temperatures.  Hence, they form an important test case for quantum simulators, and can allow the simulators to display interesting new physical phenomena.

When $N$ ions are placed in a (harmonic) linear Paul trap~\cite{MichaelJohanning}, they form a nonuniform lattice, with increasing interparticle spacing as one moves from the center to the edge of the chain. The ions vibrate in all three spatial dimensions about these equilibrium positions~\cite{James} and are described by the phonon Hamiltonian $\mathcal{H}_{ph}=\sum_{\alpha \nu}\hbar \omega_{\alpha \nu}(a^{\dag}_{\alpha \nu} a_{\alpha \nu}+1/2)$, in which $a^{\dagger}_{\alpha \nu}$ is the phonon creation operator of the normal mode $\nu$ along the spatial direction $\alpha \in {x,y,z}$.
The $\alpha$th spatial component of the $j$th ion's displacement operator $\delta {\hat R}_{j}^{\alpha}$ is related to the $\alpha$th phonon normal mode amplitude (eigenvector of the dynamical matrix) $b_{j}^{\alpha \nu}$ and the $\alpha$th phonon operator via $\delta {\hat R}_{j}^{\alpha}=\sum_{\nu} b_{j}^{\alpha \nu}[a_{\alpha \nu}+a^{\dag}_{\alpha \nu}]\sqrt{\hbar/2M\omega_{\alpha \nu}}$ with $M$ the mass of the ion.

The two clock states used to describe the quantum spins are split by an energy difference $\hbar\omega_0$.  A spin-dependent force on the ions is generated by using light to couple these states to a third state via stimulated Raman transitions~\cite{PJLee}. The light that couples to this state is detuned by a small frequency $\mu$ called the detuning.  The rotating wave approximation is used to map this three-state system onto a two-state system in the regime where $\omega_0\gg \mu$ (these differ by at least 3 orders of magnitude in typical experiments). The two-level system is then represented by an effective laser-ion Hamiltonian~\cite{Peter,PJLee,Wineland}
\begin{equation}
\mathcal{H}_{LI}(t)= -\hbar \sum_{j=1}^{N}\Omega_{j}(\delta {\bf k}\cdot\delta {\bf \hat R}_{j})\sigma_{j}^{x} \sin(\mu t)
\end{equation}
where $\delta {\bf k}$ is the wavevector difference of the two interfering laser beams that generate the Raman coupling. Here
the Lamb-Dicke limit $\delta k |\delta {\bf\hat R}_{j}(t)|\ll 1$ is assumed so that the expansion for $\exp{(-i\delta {\bf k} \cdot \delta {\bf \hat  R}_{j})}\approx 1-i\delta {\bf k} \cdot \delta {\bf \hat  R}_{j}$ is taken. The Roman index $j$ denotes the different ions,   $\Omega_{j}$ is the Rabi frequency due to the Raman coupling at the $j$th ion site, and the spin operator $\sigma_{j}^{x}$ is the Pauli spin matrix in the $x$-direction. While one can couple to any direction by tuning the phase difference of the Raman beams, for simplicity, and without loss of generality, we assume we have the lasers tuned to create a coupling to the spin component in the $x$-direction only here.
For example, we consider the ions crystallized along the $z$-axis of the trap, so the $x$- and $y$-directions correspond to transverse phonons. Then coupling the Raman lasers in the transverse direction minimizes effects of ion heating and allows for an identical spin axis for each ion~\cite{Duan}.

In experiments, one uses adiabatic quantum state evolution to evolve the ground state from an easily prepared initial state to the desired complex quantum state that will be studied. For spin models generated in an ion trap, it is easy to create a fully polarized ferromagnetic state via optical pumping, and then use a spin rotation to orient the ferromagnetic state in any direction.  Then, if one introduces a Hamiltonian with a magnetic field in the direction of the polarized state, it is in the ground state of the system. By slowly reducing the magnitude of the field and turning on the spin Hamiltonian of interest, we can reach the ground state of the Hamiltonian of interest (even in zero field).  Hence, we need to add an additional
Zeeman term  $\mathcal{H}_{B}(t)=\sum_{j=1}^{N}{\bf B}(t)\cdot{{\bf \hat \sigma}_{j}}$  with a time-dependent effective magnetic field ${\bf B}(t)$ (the coupling is to the different Pauli spin matrices). For example, a magnetic field in the $y$-direction is made by  driving a resonant radio-frequency field
with frequency $\omega_{0}$ between the two hyperfine states to implement the spin flips~\cite{Wineland}.
The full Hamiltonian is then $\mathcal{H}(t)=\mathcal{H}_{ph}+\mathcal{H}_{LI}(t)+\mathcal{H}_{B}(t)$.

Quantum dynamics for a time-dependent Hamiltonian are found by calculating the evolution operator as a time-ordered product
$U(t,t_0)=\mathcal{T}_t \exp[-i\int_{t_0}^t dt^\prime \mathcal{H}(t^\prime)/\hbar]$ and operating it on the initial quantum state $|\psi(t_0)\rangle$.  For the adiabatic evolution of the ground state, we start our system in a state with the spins aligned along the magnetic field and no phonons at time $t_0$: $|\psi(t_0)\rangle=|0\rangle_{ph}\otimes {|\uparrow_y \uparrow_y\ldots\uparrow_y\rangle}$.  In time-dependent perturbation theory, one rewrites the evolution operator in the interaction picture with respect to the time-independent part of the Hamiltonian.  This procedure produces a factorized evolution operator
\begin{equation}
U(t,t_0)=e^{-\frac{i}{\hbar} \mathcal{H}_{ph}(t-t_{0})} U_{I}(t,t_{0})
\label{eq: interaction_picture}
\end{equation}
which is the first step in our factorization procedure (the first factor on the left is called the phonon evolution operator).  The second factor on the right is the evolution operator in the interaction picture which satisfies an equation of motion given by $i \hbar {\partial U_I(t,t_{0})}{/\partial t}=[V_I(t)+\mathcal{H}_B(t)]U_I(t,t_{0})$, with $V_I(t)=\exp[i \mathcal{H}_{ph} (t-t_0)/\hbar ]\mathcal{H}_{LI}(t)\exp[-i \mathcal{H}_{ph}(t-t_{0})/\hbar]$  since $[\mathcal{H}_B(t),\mathcal{H}_{ph}]=0$. The only difference between $\mathcal{H}_{LI}(t)$ and $V_I(t)$ is that the phonon operators are replaced by their interaction picture values: $a_{\alpha\nu}\rightarrow a_{\alpha\nu}\exp [-i\omega_{\alpha\nu}(t-t_0)]$ and $a_{\alpha\nu}^\dagger
\rightarrow a_{\alpha\nu}^\dagger\exp [i\omega_{\alpha\nu}(t-t_0)]$.

We now work to factorize the evolution operator further.  Motivated by the classic problem on the driven harmonic oscillator in Landau and Lifshitz's \textit{Quantum Mechanics}~\cite{landau} and similar discussions in Gottfried's book~\cite{gottfried}, we factorize the interaction picture evolution operator further $U_{I}(t,t_{0})=\exp[-i W_{I}(t)/\hbar]\bar{U}(t,t_0)$ with $W_I(t)$ defined by $W_I(t)=\int_{t_0}^t dt^\prime V_I(t^\prime)$ (we call the factor on the right the phonon-spin evolution operator).  One important result we use is that the multiple commutator satisfies $[[W_I(t),V_I(t^\prime)],V_I(t^{\prime\prime})]=0$, which greatly simplifies the analysis below.

The equation of motion for the spin evolution operator $\bar U(t,t_{0})$ satisfies
\begin{equation}
i\hbar \frac{\partial}{\partial t}{\bar U}(t, t_{0})={\bar \mathcal{H}}(t){\bar U}(t, t_{0}),
\end{equation}
in which the operator ${\bar \mathcal{H}}(t)$ is given by the expression
\begin{equation}
{\bar \mathcal{H}}(t)=e^{\frac{i}{\hbar}W_{I}(t)}[-i\hbar \partial_{t}+\mathcal{H}_{B}(t)+V_{I}(t)]e^{-\frac{i}{\hbar}W_{I}(t)}.
\end{equation}
The operator ${\bar \mathcal{H}}(t)$ can then be expanded order by order as
\begin{eqnarray}
\label{eq:main}
e^{\frac{i}{\hbar}W_{I}(t)}V_{I}(t)e^{-\frac{i}{\hbar}W_{I}(t)}&=& V_{I}(t)+
\frac{i}{\hbar}\left[W_{I}(t), V_{I}(t)\right] \\
e^{\frac{i}{\hbar}W_{I}(t)}H_{B}(t)e^{-\frac{i}{\hbar}W_{I}(t)} &=& \sum_{j=1}^{N}\Big\{ {\bf B}(t)\cdot\hat \sigma_{j}\\ &+&\frac{i}{\hbar}
[W_{I}(t),{\bf B}(t)\cdot\hat \sigma_{j}]
+ \ldots \Big\}, \nonumber \\
e^{\frac{i}{\hbar}W_{I}(t)}i\hbar\partial_{t}e^{-\frac{i}{\hbar}W_{I}(t)}&=& {V_{I}(t)}+\frac{1}{2}\frac{i}{\hbar}[W_{I}(t),V_{I}(t)],
\end{eqnarray}
where we used the facts that $\partial{W_{I}}(t)/\partial t=V_{I}(t)$ and that $[[W_I(t),V_I(t^\prime)],V_I(t^{\prime\prime})]=0$.


Explicit calculations then yield
\begin{eqnarray}
V_{I}(t) &=& - \sum_{j=1}^{N}\sum_{\nu=1}^{N}\sum_{\alpha}\hbar\Omega_{j}\eta_{\alpha \nu} b_{j}^{\alpha \nu}(a_{\alpha \nu}e^{-i\omega_{\alpha \nu}t}+
a^{\dag}_{\alpha \nu} e^{i\omega_{\alpha \nu}t})\nonumber \\
&\times&\sin(\mu t)\sigma_{j}^{x},
 \\
W_{I}(t)&=&-\sum_{j=1}^{N}\sum_{\nu=1}^{N}\sum_{\alpha}\frac{\hbar\Omega_{j}\eta_{\alpha \nu}b_{j}^{\alpha \nu}}{\omega_{\alpha \nu}^{2}-\mu^2}\sigma_{j}^{x}\\
&\times&
\Big \{ \Big [ e^{-i\omega_{\alpha \nu}t}(i\omega_{\alpha \nu} \sin{\mu t}+\mu\cos{\mu t})\nonumber \\
&-&  e^{-i\omega_{\alpha \nu}t_{0}}(i\omega_{\alpha \nu} \sin{\mu t_{0}}+\mu\cos{\mu t_{0}}) \Big ] a_{\alpha \nu} +  h. c. \Big \},
 \nonumber \\
\end{eqnarray}
in which $\eta_{\alpha \nu}=\delta_{\alpha x}\delta k^{\alpha}\sqrt{\hbar/2M\omega_{\alpha \nu}}$ is the Lamb-Dicke parameter for the phonon mode $\alpha \nu$ with the $\alpha$th component of the laser momentum  $\delta k^{\alpha}$.

We choose the magnetic field in the $y$-direction. Then the term $\frac{i}{\hbar}[W_{I}(t'),H_{B}(t')]$ is nonzero and proportional to spin operators in the $z$-direction times phonon operators.  But it is also proportional to the Lamb-Dicke parameter $\eta_{\alpha \nu} \ll 1$ which is small in current experiments. Hence, to lowest-order in the Lamb-Dicke regime, we drop all commutator terms with the magnetic-field piece of the Hamiltonian (recall the magnetic field is a similar magnitude to the spin exchange parameters, so the additional Lamb-Dicke parameter multiplying the magnetic field makes it a less important term). The equation of motion can now be directly integrated.
The spin evolution operator ${\bar U}(t,t_{0})$ becomes
\begin{widetext}
\begin{equation}
{\bar U}(t,t_{0})  \approx  \mathcal{T}_{t}\exp\left[-\frac{i}{\hbar}\int_{t_{0}}^{t}dt' \left(\sum_{j,j'=1}^N J_{jj'}(t')\sigma_{j}^{x}\sigma_{j'}^{x}+
B(t')\sum_{j=1}^N\sigma_{j}^{y} \right)\right],
\end{equation}
\end{widetext}
which is the third factor for the evolution operator of the Ising model in a transverse field.  The spin exchange terms  $J_{jj'}(t)=J_{jj'}^{0}+\Delta J_{jj'}(t)$ arise from the $i[W_I(t),V_I(t)]/2\hbar$ commutator and include a time-independent exchange interaction between two ions $J_{jj'}^{0}$  and a time-dependent exchange interaction $\Delta J_{jj'}(t)$. The time-independent term determines the effective spin-spin Hamiltonian that is being simulated, while the time-dependent terms can be thought of as diabatic corrections, which are often small in current experimental set-ups, but need not be neglected. For simplicity, we set the initial time $t_{0} = 0$. Then the expression for the spin exchange interaction $J_{jj'}(t)$ is
\begin{eqnarray}
 J_{jj'}(t)&=&\frac{\hbar}{2}\sum_{\nu=1}^{N}\frac{\Omega_{j}\Omega_{j'}\eta_{x \nu}^{2}b_{j}^{x \nu}b_{j'}^{x \nu}}{\mu^2-\omega_{x \nu}^2}\\
 &\times&[\omega_{x\nu}-\omega_{x\nu}\cos{2\mu t}-2\mu\sin{\omega_{x\nu}t}\sin{\mu t}],\nonumber
 \end{eqnarray}
which can be antiferromagnetic $J_{jj'}(t) > 0$ or ferromagnetic $J_{jj'}(t) < 0$ depending on the laser detuning $\mu$ and the detailed phonon mode properties $\omega_{x\nu}$, $b_{j}^{x\nu}$, and $b_{j'}^{x\nu}$.

We need to evaluate one more identity before arriving at our main result. We further factorize the entangled phonon-spin evolution operator $\exp [-i W_{I}(t)/\hbar ]$ into
the product $\exp [-i\sum_{\nu}{\Gamma}_{x \nu}^{*}(t)a^{\dag}_{x \nu}]$ $\exp[-i\sum_{\nu} \Gamma_{x \nu}(t)a_{x \nu}]\exp[-\sum_{\nu}\Gamma_{x \nu}(t){\Gamma}_{x \nu}^{*} (t)/2]$
with the spin operator defined to be ${\Gamma}_{x \nu}^{*}(t)=\sum_{j}\gamma_{j}^{x\nu}(t)\sigma_{j}^{x}$
and its complex conjugate is $\Gamma_{x \nu}(t)$, in which the function $\gamma_{j}^{x\nu}(t)$ satisfies
$\gamma_{j}^{x\nu}(t)=\Omega_{j}\eta_{x \nu}b_{j}^{x\nu}/(\mu^2-\omega_{x \nu}^{2})\times[\exp\{-i\omega_{x \nu}t\}(i\omega_{x \nu}\sin\mu t+\mu \cos\mu t)-\mu]$.

We are now ready to show our main result that phonons have no effect on the probabilities to observe any of the $2^{N}$ product states with a quantization axis along the Ising axis $|\beta \rangle = |\uparrow_{x} \text{or} \downarrow_{x}\rangle \otimes |\uparrow_{x} \text{or} \downarrow_{x}\rangle\otimes \ldots$ for the $N$ ionic spins. Using the fundamental axiom of quantum mechanics, the probability $P_\beta(t)$ to observe a product spin state $|\beta \rangle$ starting initially from the phonon ground state (for all phonon modes) $|0\rangle$ and not measuring any of the final phonon states involves the trace over all possible final phonon configurations
\begin{widetext}
\begin{equation}
\label{eq:prob}
P_\beta(t)=\sum_{n_{x1}=0}^{\infty}\cdots\sum_{n_{x\nu}=0}^{\infty}\cdots\sum_{n_{x N}=0}^{\infty}|\langle\beta| \otimes {}_{ph}\langle n_{x 1},\ldots n_{x \nu},\ldots n_{x N}|  U(t,t_{0})|0,0,\ldots,0\rangle_{ph} \otimes |\Phi(0)\rangle|^{2}
\end{equation}
\end{widetext}
where $|\Phi(0)\rangle$ is any initial spin state (it need not be a product state). 
Note that since the pure phonon factor of the evolution operator $\exp[-i\mathcal{H}_{ph}t/\hbar ]$ is a phase factor, it has no effect on the probabilities when evaluated in the phonon number operator basis, so we can drop that factor. Next, the term $\exp[-i\sum_{\nu} \Gamma_{x \nu}(t)a_{x \nu}]$ gives 1 when operating on the phonon vacuum to the right, so it can be dropped. We are thus left with three factors in the evolution operator.  One involves exponentials of the phonon creation operator multiplied by spin operators, one involves products of spin operators that resulted from the factorization of the coupled phonon-spin evolution operator factor, and one is the pure spin evolution factor $\bar U$. The two factors that appear on the left involve only Ising spin operators, and hence the product state basis is an eigenbasis for those operators.  This fact allows us to directly evaluate the expression in Eq.~(\ref{eq:prob}). We expand the evolution of the initial state at time $t$ in terms of the product-state basis
$|\Phi(t)\rangle = {\bar U}(t,t_{0}=0)|\Phi_{S}^{0}\rangle=\sum_{\beta^\prime} c_{\beta^\prime}(t)|{\beta^\prime}\rangle$ with $|\beta^\prime\rangle$ denoting each of the $2^N$ product state basis vectors and $c_{\beta^\prime}(t)$ is a number.
Using the fact that the product states satisfy the eigenvalue equation $\sigma_{j}^{x}|\beta^\prime\rangle = m_{j\beta^\prime}^{x}|\beta^\prime\rangle$ with eigenvalues $m_{j\beta^\prime}^{x}=+1 $ (for $|\uparrow_{x}\rangle$) or $-1$ (for $|\downarrow_{x}\rangle$), we arrive at the expression for the probability
\begin{widetext}
\begin{equation}
P_{\beta}(t) = |c_{\beta}(t)|^2 \exp{\left [ -\sum_{\nu j j'}\gamma^{x\nu}_{j}(t)\gamma^{x\nu *}_{j'}(t)m_{j\beta}^{x}m_{j'\beta}^{x}\right ]}
 \sum_{n_{x1}=0}^{\infty}\cdots\sum_{n_{x N}=0}^{\infty}
\prod_{\nu=1}^N\left [ \frac{1}{n_{x\nu}!}\left \{\sum_{jj'}\gamma_{j}^{x\nu}(t)\gamma_{xj'}^{x\nu *}(t)
m_{j\beta}^{x}  m_{j'\beta}^{x} \right\}^{n_{x\nu}}\right ] .
\end{equation}
\end{widetext}
We used the matrix element
\begin{eqnarray}
&{}_{ph}\langle n_{x 1},\ldots, n_{x \nu}, \ldots n_{x N}|e^{-i\sum_{\nu}{\Gamma}_{x \nu}^{*}(t)a^{\dag}_{x \nu}}|0,\ldots 0\rangle_{ph} \\
&=\frac{{(-i)}^{n_{x1}+...n_{x\nu}+..n_{xN}}}{\sqrt{n_{x1}!..n_{x\nu}!..n_{xN}!}}{\Gamma}_{x 1}^{*}(t)^{n_{x1}}...{\Gamma}_{x \nu}^{*}(t)^{n_{x\nu}}
...{\Gamma}_{x N}^{*}(t)^{n_{xN}}\nonumber
\end{eqnarray}
in the derivation. The summations become exponentials, which exactly cancel the exponential term and finally yield $P_{\beta}(t)=|c_{\beta}(t)|^2$, which is what we would have found if we evaluated the evolution of the spins using just the spin evolution operator $\bar U$.
Hence, \textit{phonons have no observable effects on the probability of product states for the transverse-field quantum Ising model}.
If we do not measure the probability of product states, then the terms from the coupled phonon-spin evolution operator remain spin operators, and one can show that the probabilities are changed by the phonons by simply evaluating the result in Eq.~(\ref{eq:prob}) directly.  In other words, it is because the $\Gamma$ operators are diagonal in the product space basis along the Ising axis that allows us to disentangle the phonon and spin dynamics.  In cases where this cannot be done, we expect the phonon and spin dynamics to generically remain entangled. In particular, when one measures a typical entanglement witness operator, one measures the probabilities for the state to be projected onto product states that do not lie along the Ising axis.  In this case, the phonons and spins will remain entangled, and the phonons will affect the expectation value of the witness operator.

The simplest way to set up XY or Heisenberg model is to introduce additional laser beams with perpendicular $\delta {\bf k}$ values so one gets pure spin-spin interactions along each of the different coordinate axes (if one cannot introduce perpendicular beams, then other methods can be tried, but they require strategies to reduce unwanted cross terms like $\sigma^x\sigma^y$).  For the XY or Heisenberg model, we generate additional entangled spin-phonon evolution operator factors for the different coordinate directions ($x$ and $y$ for the XY model and all for the Heisenberg).  But, since one cannot find a basis of quantum states that simultaneously diagonalizes the Pauli spin operators in more than one coordinate direction, the above derivation cannot be completed, and so the phonons will generically modify the probabilities for the spin states.

In conclusion, we have explicitly shown that phonons do not affect the probability of observing spin product states in ion trap simulators of the transverse-field Ising model to lowest order in the Lamb-Dicke parameter.  This implies that simulations of the transverse-field Ising model have the phonon motion and the spin evolution disentangled, and hence will be much cleaner than simulations of the XY or Heisenberg model.  Our results do not depend on the detuning lying far enough from resonance, unless the proximity to resonance moves the system out of the Lamb-Dicke regime so higher-order terms need to be included, in which case phonon entanglement effects are likely to affect the results.

\section{Acknowledgements}
We acknowledge useful discussions with Kihwan Kim, Chris Monroe, and L.-M Duan.
This work was supported under ARO grant number W911NF0710576 with funds from the DARPA OLE Program.

\end{document}